\documentclass{elsart}
\usepackage{graphicx,amsmath,bm}

\begin{document}

\begin{frontmatter}

\title{An isobar model for $\bm{\eta}$ photo- and electroproduction
       on the nucleon}
\author{Wen-Tai Chiang and Shin Nan Yang}
\address{Department of Physics, National Taiwan University,
              Taipei 10617, Taiwan}
\author{Lothar Tiator and Dieter Drechsel}
\address{Institut f\"ur Kernphysik, Universit\"at Mainz,
                55099 Mainz, Germany}

\begin{abstract}
Eta photo- and electroproduction on the nucleon is studied using an isobar
model. The model contains Born terms, and contributions from vector meson
exchanges and nucleon resonances. Our results are compared with recent eta
photoproduction data for differential and total cross sections, beam
asymmetry, and target asymmetry, as well as electroproduction data. Besides
the dominant $S_{11}(1535)$ resonance, we show that the second $S_{11}$
resonance, $S_{11}(1650)$, is also necessary to be included in order to
extract $S_{11}(1535)$ resonance parameters properly. In addition, the beam
asymmetry data allow us to extract very small ($<0.1\%$) $N^* \rightarrow
\eta N$ decay branching ratios of the $D_{13}(1520)$ and $F_{15}(1680)$
resonances because of the overwhelming $s$-wave dominance. This model
(ETA-MAID) is implemented as a part of the MAID program~\footnote{The MAID
program can be accessed from the webpage:
\texttt{http://www.kph.uni-mainz.de/MAID/maid.html.}}.
\end{abstract}


\end{frontmatter}

\section{Introduction} \label{sec:Intro}
Eta photo- and electroproduction on the nucleon, $\gamma^* N \rightarrow
\eta N$, provide an alternative tool to study $N^*$ besides $\pi N$
scattering and pion photoproduction. There are fewer resonances involved
since the $\eta N$ state couples to nucleon resonances with isospin $I =
1/2$ only. Therefore, this process is cleaner and more selective to
distinguish certain resonances than other processes, e.g., pion
photoproduction. This provides opportunities to access less studied
resonances and possibly the ``missing resonances''.

During the last decade eta photo- and electroproduction has been studied in
very different frameworks. Common in most approaches is the dominance of
the $S_{11}(1535)$ resonance, which seems to be the natural mechanism to
describe the almost constant angular distributions of the differential
cross sections from threshold at $W=1487\;\mathrm{MeV}$ up to energies
around $W=1700\;\mathrm{MeV}$. Early coupled channel approaches predicted
eta photoproduction on the basis of elastic and inelastic pion nucleon
scattering and pion photoproduction amplitudes~\cite{Bennhold:1991kj}, more
recent developments included the eta photoproduction data in their
fits~\cite{Sauermann:1995pu,Feuster:1998cj,Waluyo:2001}. Field theoretical
Lagrangian
models~\cite{Benmerrouche:1995uc,Mukhopadhyay:1998hn,Davidson:2000in} and
applications of chiral perturbation
theory~\cite{Borasoy:2000pj,Borasoy:2000kh} derive covariant eta production
amplitudes from Born terms, vector meson exchange, and resonance
excitations. In these approaches the spin 3/2 resonances such as
$D_{13}(1520)$ are treated in a Rarita-Schwinger formalism but higher spin
resonances have not yet been introduced. Relativistic quark models have
been successful in describing the general behavior of the cross
sections~\cite{Li:1995sia,Li:1997gda}, and in a very recent approach a very
good fit of the existing data has been obtained by introducing a third
$S_{11}$ resonance around 1700 MeV~\cite{Saghai:2001yd} in addition to the
well-known resonances $S_{11}(1535)$ and $S_{11}(1650)$. The role of
polarization observables has been studied in an isobar
model~\cite{Knochlein:1995qz} based on early experiments from
Bonn~\cite{Schoch:1995} and Mainz~\cite{Krusche:1995nv}.

Apart from the $S_{11}(1535)$ dominated approaches, alternative ways have
been discussed within the Lee Model as well as in chiral meson-baryon
Lagrangian theory. In these approaches the $s$-wave production is either
described by a nonresonant background~\cite{Denschlag:1998qn} or by a
molecule like $K\Sigma$ intermediate state. As was pointed out by
H\"ohler~\cite{Hohler:1992,Hohler:1993}, the existence of the
$S_{11}(1535)$ cannot be unambiguously proven by the standard technique of
pion nucleon speed-plots. A pole position of $1505-85i$ as given by
PDG~\cite{PDG:2000} is too close to threshold with a width larger than the
distance from threshold in order to show a peak in the speed-plot. The only
signature that appears is the sharp spike due to the $\eta$ cusp. On the
other hand the resonance is strongly supported by the quark model and even
the large $\eta N$ branching ratio of almost 50\% can be understood by
color hyperfine mixing~\cite{Koniuk:1980vy}. The same resonance also makes
a significant contribution in pion photo- and electroproduction. There,
however, it usually causes the problem of a photon resonance coupling
$A_{1/2}$ of only $0.5-0.6$ times the value obtained in eta photoproduction
analyses~\cite{Benmerrouche:1995uc,Krusche:1997jj}. In order to shed more
light on the underlying nature of the very pronounced $s$-wave amplitude, a
precise study of the $Q^2$ dependence of both transverse and longitudinal
couplings is needed for both pion and eta electroproduction. In addition
the same mechanism also has to describe the eta production off neutrons
that has been investigated in coherent and incoherent eta photoproduction
experiments on the deuteron. Only by introducing a strong background in the
$s$-wave amplitude both reactions could be described
simultaneously~\cite{Ritz:2001ag}.

The situation of higher resonances is somewhat different. Their existence
is well established but the decay in the $\eta N$ channel is mostly very
vague. Branching ratios listed in PDG92 were later removed for being too
uncertain. Only very recently, due to precise photon asymmetry data from
GRAAL, a branching ratio of $(0.08 \pm 0.01)\%$ could be determined for the
$D_{13}(1520)$ resonance~\cite{Tiator:1999gr}. This result has also been
confirmed by other analyses of the same data set. In these analyses another
discrepancy in comparison to pion photoproduction was observed yielding a
much smaller $A_{3/2}/A_{1/2}$ ratio for the $D_{13}(1520)$ resonance
\cite{Mukhopadhyay:1998hn,Tiator:1999gr,Workman:2000wt}.

Unlike pion production for which Born terms give large background
contributions to all partial waves due to the large pion nucleon coupling
constant, in $\eta$ production the coupling of the eta meson to the nucleon
is very small. Already in the SU(3) limit the coupling $g_{\eta NN}^2 /
4\pi = 0.8 - 1.9$ is much smaller than for pions ($g_{\pi NN}^2 / 4\pi =
14.3$), but in an analysis of the angular distributions of eta
photoproduction an even smaller value of $g_{\eta NN}^2 / 4\pi = 0.4 \pm
0.2$ was determined~\cite{Tiator:1994et}. Such a small value was later also
explained within a chiral Lagrangian
approach~\cite{Kirchbach:1996kw,Neumeier:2000fb}, and in a very recent fit
within a chiral constituent quark model a value of only $0.04$ has been
obtained~\cite{Saghai:2001yd}. Therefore, the only sizeable background
contribution remains in the $t$-channel vector meson amplitudes, mainly due
to $\rho^0$ exchange.

The aim of this paper is to extend an earlier version of an isobar model by
Kn\"{o}chlein {\it et al.}~\cite{Knochlein:1995qz} and to continue the work
on the unitary isobar model MAID~\cite{Drechsel:1998hk}. In the same way as
in MAID for pions we will describe the eta photo- and electroproduction in
terms of nucleon Born terms, vector meson exchange contributions and
resonance excitations parameterized with Breit-Wigner shapes directly
connected to the conventional resonance parameters listed in the particle
data tables: masses, widths, branching ratios and photon couplings. By use
of the world data on photo- and electroproduction, all free or uncertain
parameters of the model will be fixed by least squares fitting. These data
are the total and differential photoproduction cross sections of
MAMI~\cite{Krusche:1995nv} and GRAAL~\cite{Renard:2000iv}, the photon
asymmetry of GRAAL~\cite{Ajaka:1998zi} and the electroproduction cross
sections of JLab~\cite{Armstrong:1998wg,Thompson:2000by}.

In Section~2, we give the general formalism of photo- and
electroproduction, and describe the model ingredients and resonance
parameterization in Section~3. Our fitting results and predictions for
future experiments are given in Section~4 followed by a summary and
conclusions.

\section{Formalism} \label{sec:Form}


Consider $\eta$ electromagnetic production on the nucleon. This reaction
includes (1) photoproduction
\begin{equation} \label{eq:phoprod}
  \gamma\,(k) + N\,(p_i) \longrightarrow \eta\,(q) + N\,(p_f) \,,
\end{equation}
and (2) electroproduction
\begin{equation} \label{eq:elecprod}
  e\,(k_i) + N\,(p_i) \longrightarrow e'\,(k_f) + \eta\,(q) + N\,(p_f) \,,
\end{equation}
where the four-momentum for each particle is indicated in the parentheses.
The four-momentum of the virtual photon exchanged in electroproduction is
given by $k = k_i - k_f$, with $k^2 = 0$ for photoproduction. The
four-momentum square of the virtual photon is negative; therefore, in this
context we use a positive quantity $Q^2 \equiv -k^2 = \bm{k}^2 - \omega^2$
to describe form factors and structure functions.

In the $\gamma N / \eta N$ center-of-mass (c.m.) frame, the momenta of the
initial $\gamma N$ and final $\eta N$ states can be expressed in terms of
the total c.m. energy $W = \sqrt{s}$ and $k^2$,
\begin{eqnarray}
  |\bm{k}| = |\bm{p_i\,}| &=& \frac{1}{2W}
  \sqrt{ \bigl((W+M_i\,)^2+Q^2\bigr) \bigl((W-M_i\,)^2+Q^2\bigr) } \,,
  \nonumber\\
  |\bm{q}| = |\bm{p_f}| &=& \frac{1}{2W}
  \sqrt{ \bigl((W+M_f)^2-m^2\bigr) \bigl((W-M_f)^2-m^2\bigr) } \,.
\end{eqnarray}
In photoproduction, the relation between the photon energy $E_\gamma^{lab}$
in the lab frame and the total c.m. energy $W$ is
\begin{equation}
  E_\gamma^{lab} = \frac{W^2-M_i^2}{2M_i}\,.
\end{equation}


\begin{figure}
  \centering
  \includegraphics[width=9cm]{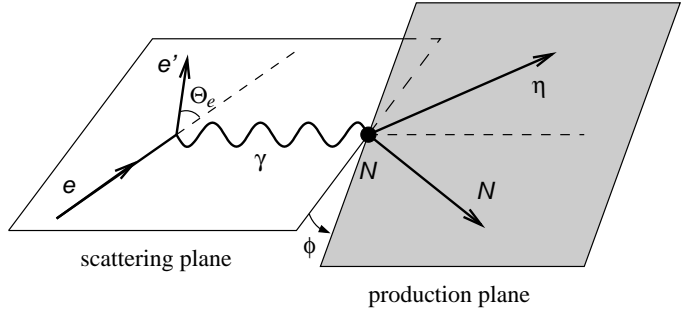}
  \caption{Kinematics in the lab frame for $\eta$ electroproduction
           $e N \rightarrow e' \eta N$}
  \label{fig:plane}
\end{figure}

Following the convention of Bjorken and Drell~\cite{Bjorken:1964a}, the
differential cross section for the electroproduction process can be written
as
\begin{eqnarray}
\d\sigma &=& \frac{m_e M_i}{\sqrt{(k_i \cdot p_i)^2 - k_i^2\,p_i^2}}
  \frac{d^3 k_f}{(2\pi)^3} \frac{m_e}{\varepsilon_f}
  \frac{d^3 q}{(2\pi)^3} \frac{1}{2\omega_\eta}
  \frac{d^3 p_f}{(2\pi)^3} \frac{M_f}{E_f}
\nonumber \\
  & & \times (2\pi)^4 \,
  \delta^{(4)}(k + p_i - q - p_f) \,
  \bigl| \langle\,p_f\,q\,|\,J^{\mu}\,|\,p_i\,\rangle \frac{1}{\;k^2\,}
         \langle\,k_f\,|\,j_{\mu}\,|\,k_i\,\rangle  \bigr|^2 \; ,
\end{eqnarray}
where $j^{\mu}$ and $J^{\mu}$ denote the electromagnetic currents of the
electron and the hadronic system, respectively.

The transverse polarization parameter of the virtual photon
\begin{equation}
 \varepsilon = \left(1 + 2 \frac{\,\bm{k}^2}{\,Q^2}\tan^2\frac{\Theta_e}{2}
               \right)^{-1} \,,
\end{equation}
is invariant under collinear transformations, and $\bm{k}$ and $\Theta_e$
may be expressed in the lab or c.m. frame. By choosing the energies of the
initial and final electrons and the scattering angle $\Theta_e$ (see
Fig.~\ref{fig:plane}), we can fix the momentum transfer $Q^2$ and the
polarization parameter $\varepsilon$ of the virtual photon.

The five-fold differential cross section for electroproduction can be
expressed as~\cite{Donnachie:1978fm,Amaldi:1979vh,Drechsel:1994}
\begin{equation} \label{eq:5fdxs}
 \frac{\d\sigma}{\d\Omega_f\, \d\varepsilon_f\, \d\Omega}
 = \Gamma\, \frac{\d\sigma_v}{\d\Omega} \,,
\end{equation}
with the flux of the virtual photon field given by
\begin{equation}
 \Gamma = \frac{\alpha}{2 \pi^2}\, \frac{\varepsilon_f}{\varepsilon_i}\,
          K\, \frac{1}{1-\varepsilon} \,,
\end{equation}
where $\varepsilon_i$ and $\varepsilon_f$ are the initial and final
electron energies. In this expression $K \equiv (W^2 - M_i^2) / 2 M_i $
denotes the ``photon equivalent energy'', the laboratory energy necessary
for a real photon to excite a hadronic system with c.m. energy $W$. It is
useful to express the angular distribution of the eta mesons in the c.m.
frame of the final hadronic states, particularly for the use of multipole
decompositions. Therefore, the virtual photon cross section $\d\sigma_v /
\d\Omega$ should be evaluated in the c.m. frame, while the five-fold
differential cross section in the Eq.~(\ref{eq:5fdxs}) is interpreted with
the flux factor in the lab frame. For the rest of the paper we only use
c.m. variables.

For an unpolarized target and without recoil polarization, the virtual
photon differential cross section is
\begin{eqnarray} \label{eq:virdxs}
 \frac{\d\sigma_v}{\d\Omega}
 &=& \frac{\d\sigma_T}{\d\Omega}
     +  \varepsilon\, \frac{\d\sigma_L}{\d\Omega}
     +  \sqrt{2 \varepsilon (1+\varepsilon)}\;
        \frac{\d\sigma_{LT}}{\d\Omega}\, \cos\phi
 \nonumber \\
 & & + \varepsilon\, \frac{\d\sigma_{TT}}{\d\Omega}\, \cos2\phi
     +  h\, \sqrt{2 \varepsilon (1-\varepsilon)}\;
       \frac{\d\sigma_{LT'}}{\d\Omega}\, \sin\phi \,.
\end{eqnarray}
Here $\phi$ is the azimuthal angle between the electron scattering plane
and the $\eta$ production plane (see Fig.~\ref{fig:plane}), and $h =
\bm{\sigma} \cdot \hat{\bm{k}}_i = \pm 1$ is the helicity of the incident
electron with longitudinal polarization. The first two terms ($\d\sigma_T$
and $\d\sigma_L$) in the RHS of Eq.~(\ref{eq:virdxs}) are referred as the
transverse and longitudinal cross sections, and do not depend on the
azimuthal angle $\phi$. The $\d\sigma_{LT}$ and $\d\sigma_{LT'}$ describe
longitudinal-transverse interferences, and the $\d\sigma_{TT}$ is a
transverse-transverse interference term. We do not use the longitudinal
polarization $\varepsilon_L$; therefore, the cross sections with
longitudinal components ($\d\sigma_L$, $\d\sigma_{LT}$ and
$\d\sigma_{LT'}$) differ from those in
Refs.~\cite{Knochlein:1995qz,Drechsel:1994}.

\begin{figure}
  \centering
  \includegraphics[width=6cm]{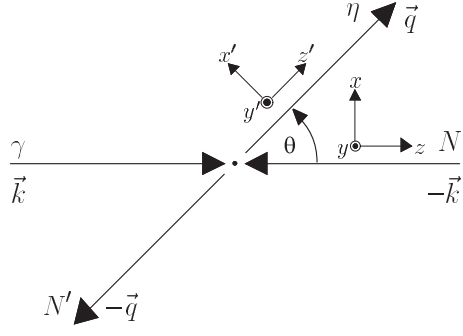}
  \caption{Frames for polarization vectors in $\eta$ photoproduction
           in the c.m. system}
  \label{fig:polvec}
\end{figure}

Three types of polarization measurements can be performed in pseudoscalar
meson production: photon beam polarization, polarization of the target
nucleon, and polarization of the recoil nucleon. Target polarization is
described in the frame $\{ x, y, z \}$ of Fig.~\ref{fig:polvec}, with the
$z$-axis pointing into the direction of the photon momentum $\hat{\bm{k}}$,
the $y$-axis perpendicular to the reaction plane, $\hat{\bm{y}} =
\hat{\bm{k}} \times \hat{\bm{q}} / \sin \theta$, and the $x$-axis given by
$\hat{\bm{x}} = \hat{\bm{y}} \times \hat{\bm{z}}$. For recoil polarization
we will use the frame $\{ x', y', z' \}$, with the $z'$-axis defined by the
outgoing $\eta$ momentum vector $\hat{\bm{q}}$, the $y'$-axis as for target
polarization and the $x'$-axis given by $\hat{\bm{x}}' = \hat{\bm{y}}'
\times \hat{\bm{z}}'$. The most general expression for a coincidence
experiment considering all three types of polarization is
\begin{eqnarray} \label{eq:dsigmafull}
 \frac{\d\sigma_v}{\d\Omega} &=&
 \frac{|\,\bm{q}\,|}{|\,\bm{k}\,|} P_{\alpha} P_{\beta} \Bigl\{
       R_T^{\beta \alpha}
     + \varepsilon\, \tfrac{Q^2}{\omega^2} R_L^{\beta\alpha}
       \nonumber \\
 & & + \,\sqrt{2\,\varepsilon (1+\varepsilon) \tfrac{Q^2}{\omega^2}}\,
       \bigl({\,}^c \! R_{LT}^{\beta\alpha} \cos \phi
     + {\,}^s \! R_{LT}^{\beta\alpha} \sin \phi \bigr)
       \nonumber \\ [0.4ex]
 & & + \,\varepsilon\,\bigl({\,}^c \! R_{TT}^{\beta\alpha} \cos 2\phi
     + {\,}^s \! R_{TT}^{\beta\alpha} \sin 2 \phi \bigr)
       \nonumber \\ [0.2ex]
 & & + \,h\,\sqrt{2\,\varepsilon (1-\varepsilon) \tfrac{Q^2}{\omega^2}}\,
       \bigl({\,}^c \! R_{LT'}^{\beta\alpha} \cos \phi
     + {\,}^s \! R_{LT'}^{\beta\alpha} \sin \phi \bigr)
       \nonumber \\
 & & + \,h\,\sqrt{1-\varepsilon^2}\,R_{TT'}^{\beta\alpha}
 \Bigr\} \,,
\end{eqnarray}
where $P_{\alpha} = (1, \bm{P})$ and $P_{\beta} = (1, \bm{P}')$. Here
$\bm{P} = (P_x, P_y, P_z)$ and $\bm{P}' = (P_{x'}, P_{y'}, P_{z'})$ denotes
the target and recoil polarization vector. The zero components $P_0 = 1$
correspond to contributions in the cross section which are present in the
polarized as well as the unpolarized case. In an experiment without target
and recoil polarization $\alpha = \beta = 0$, therefore the only remaining
contributions are $R_i^{00}$. The functions $R_i^{\beta \alpha}$ describe
the response of the hadronic system in the process. Summation over Greek
indices is implied. An additional superscript $s$ or $c$ on the left
indicates a sine or cosine dependence of the respective contribution on the
azimuthal angle. Some response functions vanish identically. See
Ref.~\cite{Knochlein:1995qz} for a systematic overview.

In photoproduction the longitudinal components vanish due to the external
factors $\sqrt{Q^2/\omega^2}$ ($\omega$ is the c.m. energy of the virtual
photon), and in the following the relevant response functions will be
divided by the (unpolarized) transverse response function $R_T^{00}$ in
order to obtain the polarization observables. The common descriptors of
these observables can be found in Table~\ref{tbl:phprod}.

\begin{table}
\caption{Polarization observables in pseudoscalar meson photoproduction.
         The entries in parentheses signify that the same polarization
         observables also appear elsewhere in the table.}
\label{tbl:phprod}%
\vspace{2mm}
\renewcommand{\arraystretch}{1.6}
\scriptsize%
\begin{tabular}{c c ccc ccc cccc}
\hline
 Photon &  & \multicolumn{3}{c}{Target} & \multicolumn{3}{c}{Recoil} &
 \multicolumn{4}{c}{Target + Recoil} \\
\hline
 & $-$ & $-$ & $-$ & $-$ & $x'$ & $y'$ & $z'$ & $x'$ & $x'$ & $z'$ & $z'$ \\
 & $-$ & $x$ & $y$ & $z$ & $-$ & $-$ & $-$ & $x$ & $z$ & $x$ & $z$ \\
\hline
 unpolarized & $\sigma_0$ & $0$ & $T$ & $0$ & $0$ & $P$ & $0$ &
               $T_{x'}$ & \!$-L_{x'}$\! & $T_{z'}$ & $L_{z'}$ \\
 linear pol. & $\ -\Sigma\ $ & $\ H\ $ & \!$(-P)$\! &
               $-G$ & $\ O_{x'}$ & \!\!$(-T)$\!\! & $\ O_{z'}$ &
               \!$(-L_{z'})$\! & \!\!$(\,T_{z'}\,)$\!\! &
               \!$(-L_{x'})$\! & \!\!$(-T_{x'})$\!\!\\
 circular pol. & $\quad 0\quad$ & $F$ & $0$ & $-E$ &
                 $-C_{x'}$ & $0$ & $-C_{z'}$ &
                 $0$ & $0$ & $0$ & $0$ \\
\hline
\end{tabular}
\end{table}

In contrast to electroproduction, there are no new independent polarization
observables accessible by triple (beam + target + recoil) polarization
measurements in photoproduction. As a consequence we classify the
differential cross sections by the three classes of double polarization
experiments:
\begin{itemize}
\item polarized photons and polarized target
\end{itemize}
\begin{eqnarray}
  \frac{\d\sigma}{\d\Omega} & = &
      \sigma_0 \bigl[\, 1 - P_{T\,} \Sigma \cos 2 \varphi
      + P_x ( - P_T H \sin 2 \varphi + P_{\odot} F )
      \nonumber \\
  & & - P_y ( - T + P_T P \cos 2 \varphi )
      - P_z ( - P_T G \sin 2 \varphi + P_{\odot} E )
      \bigr] \,,
\end{eqnarray}
\begin{itemize}
\item polarized photons and recoil polarization
\end{itemize}
\begin{eqnarray}
  \frac{\d\sigma}{\d\Omega} & = &
      \sigma_0 \bigl[\, 1 - P_{T\,} \Sigma \cos 2 \varphi
      + P_{x'} ( - P_T O_{x'} \sin 2 \varphi - P_{\odot} C_{x'} )
      \nonumber \\
  & & - P_{y'} ( - P + P_T T \cos 2 \varphi )
      - P_{z'} ( P_T O_{z'} \sin 2 \varphi + P_{\odot} C_{z'}) \bigr] \,,
\end{eqnarray}
\begin{itemize}
\item polarized target and recoil polarization
\end{itemize}
\begin{eqnarray}
  \frac{\d\sigma}{\d\Omega} & = &
      \sigma_0 \bigl[\, 1 + P_{y'} P
      + P_x ( P_{x'} T_{x'} + P_{z'} T_{z'}) \nonumber \\
  & & + P_{y} ( T + P_{y'} \Sigma )
      - P_{z} \left( P_{x'} L_{x'} - P_{z'} L_{z'} \right) \bigr] \,,
\end{eqnarray}
where $\sigma_0$ denotes the unpolarized differential cross section, $P_T$
the transverse degree of photon polarization, $P_{\odot}$ the right-handed
circular photon polarization, and $\varphi$ the angle between photon
polarization vector and reaction plane.

In conclusion, there are 16 different polarization observables for real
photon experiments. In electroproduction there are four additional
observables for the exchange of longitudinal photons and sixteen
observables due to longitudinal-transverse interference. However, there are
only six independent complex amplitudes for the electroproduction process.
This corresponds to six absolute values and five relative phases between
the CGLN amplitudes; i.e. there are only eleven independent quantities
which completely and uniquely determine the transition current $J^{\mu}$.

The aim of a so-called ``complete experiment'' is to fully determine the
current $J^{\mu}$ for the process under investigation in a given
kinematics. For the photoproduction case a minimum of eight carefully
chosen observables can determine amplitudes without any
ambiguities~\cite{Chiang:1996em}. Although it would be interesting to
pursue such a project for $\eta$ production, we will concentrate on the
more physical aspects of this process. In particular our theoretical
investigations will supply information on an adequate selection of response
functions and observables. In the case of $\eta$ production it will be of
special interest to study observables which give clear information of the
eta meson coupling to higher resonances.

We can also introduce amplitudes defined by the helicity eigenstates of the
initial and final nucleons and the photon in
photoproduction~\cite{Walker:1969xu}. The connections between the helicity
amplitudes $H_i$ and the CGLN amplitudes $\mathcal{F}_i$ are
\begin{eqnarray} \label{eq:helamp}
  H_1(\theta) &=& -\tfrac{1}{\sqrt{2}}\,\sin\theta\,\cos\tfrac{\theta}{2}\,
  \bigl( \mathcal{F}_3 + \mathcal{F}_4 \bigr)\,e^{i\phi} \,,
  \nonumber \\[0.5ex]
  H_2(\theta) &=& \sqrt{2}\,\cos\tfrac{\theta}{2}\,
  \bigl[ (\mathcal{F}_2-\mathcal{F}_1)
  + \sin^2\tfrac{\theta}{2}\,(\mathcal{F}_3-\mathcal{F}_4) \bigr] \,,
  \nonumber \\[0.5ex]
  H_3(\theta) &=& \tfrac{1}{\sqrt{2}}\,\sin\theta\,\sin\tfrac{\theta}{2}\,
  \bigl( \mathcal{F}_3 - \mathcal{F}_4 \bigr)\,e^{2i\phi} \,,
  \nonumber \\[0.5ex]
  H_4(\theta) &=& \sqrt{2}\,\sin\tfrac{\theta}{2}\,
  \bigl[ (\mathcal{F}_2+\mathcal{F}_1)
  + \cos^2\tfrac{\theta}{2}\,(\mathcal{F}_3+\mathcal{F}_4) \bigr]\,
  e^{i\phi} \,,
  \nonumber \\[0.5ex]
  H_5(\theta) &=& \cos\tfrac{\theta}{2}\,
  \bigl( \mathcal{F}_6 + \mathcal{F}_5 \bigr) \,,
  \nonumber \\[0.5ex]
  H_6(\theta) &=& \sin\tfrac{\theta}{2}\,
  \bigl( \mathcal{F}_6 - \mathcal{F}_5 \bigr)\,e^{i\phi} \,.
\end{eqnarray}

\section{Isobar model} \label{sec:Model}
The isobar model used in this work is closely related to the unitary isobar
model (MAID) developed by Drechsel {\it et al.}~\cite{Drechsel:1998hk}. The
major difference is that in MAID, which deals with pion photo- and
electroproduction, the phases of the multipole amplitudes are adjusted to
the corresponding pion-nucleon elastic scattering phases. However, in
$\eta$ production the unitarization procedure is not feasible, because the
necessary information on eta-nucleon scattering is not available.

\subsection{Background} \label{sec:Backgd}
The nonresonant background contains the usual \emph{Born terms} and
\emph{vector meson exchange} contributions. It is obtained by evaluating
the Feynman diagrams derived from an effective Lagrangian. For the
electromagnetic $\gamma N N$ vertex the structure is well understood,
\begin{equation} \label{eq:LgNN}
 \mathcal{L}_{\gamma NN} = -e\,\bar\psi \left[\gamma_{\mu} A^{\mu}
 F_{1}^{p,n}(Q^{2}) + \frac{\sigma_{\mu\nu}}{2m_{N}}(\partial^{\mu}A^{\nu})
 F_{2}^{p,n}(Q^{2})\right]\psi \,,
\end{equation}
with $A^{\mu}$ the electromagnetic vector potential, and $\psi$ the nucleon
field operators. In Eq.~(\ref{eq:LgNN}) we have included proton
($F_{1,2}^{p}$) and neutron ($F_{1,2}^{n}$) electromagnetic form factors
with explicit $Q^2$ dependence. In the case of real photons the form
factors are normalized to $F_1^p(0)=1$, $F_1^n(0)=0$,
$F_2^p(0)=\kappa_p=1.79$ and $F_2^n(0)=\kappa_n=-1.97$. For virtual photons
the nucleon form factors are expressed in terms of the Sachs form factors
by the standard dipole form, $F_D(Q^2) = (1+Q^2/0.71(\mbox{GeV/c})^2)^{-2}$

The description of the hadronic $\eta N N$ vertex is more sophisticated in
the pseudoscalar meson electroproduction. There are two possibilities for
constructing the interaction Lagrangian, namely, the pseudoscalar (PS)
coupling,
\begin{equation} \label{eq:LPS}
 \mathcal{L}_{\eta N N}^{PS} = - i\, g_{\eta N N}\,
 \bar{\psi}\, \gamma_5\, \psi\, \phi_\eta \,,
\end{equation}
and the pseudovector (PV) coupling,
\begin{equation} \label{eq:LPV}
 \mathcal{L}_{\eta N N}^{PV} = \frac{f_{\eta N N}}{m_\eta}\,
 \bar{\psi}\, \gamma_5\, \gamma_\mu\, \psi\, \partial^\mu \phi_\eta \,,
\end{equation}
where the two types of coupling are related by $f_{\eta N N}/m_\eta=g_{\eta
N N}/2m_N$. In contrast to the $\pi N$ interaction, where PS coupling is
ruled out by chiral symmetry, both couplings are allowed for the $\eta N$
interaction and we have chosen the PS coupling in accordance with
Ref.~\cite{Tiator:1994et}.

Using the effective Lagrangians Eqs.~(\ref{eq:LgNN})-(\ref{eq:LPV}), we
construct the usual \emph{Born terms}. The other part of background is
\emph{vector meson exchange} contributions. The effective Lagrangians for
the vector meson exchange vertices are
\begin{eqnarray}
 {\mathcal{L}}_{\gamma \eta V} & = & \frac{e \lambda_V}{m_{\eta}}\,
 \varepsilon_{\mu \nu \rho \sigma}\,(\partial^{\mu} A^{\nu})\,\phi_{\eta}\,
 (\partial^{\rho} V^{\sigma})\, F_V^{em}(Q^2) \,, \label{eq:LgeV}
 \\ [1ex]%
 {\mathcal{L}}_{V N N} & = & {\bar{\psi}} \left( g_v \gamma_{\mu} +
 \frac{g_t}{2 m_N}\, \sigma_{\mu \nu}
 \partial^{\nu} \right) V^{\mu} \psi \,, \label{eq:LVNN}
\end{eqnarray}

The parameters for the $\rho$ and $\omega$ mesons in this model are listed
in Table \ref{tbl:vecmes}. The electromagnetic couplings of the vector
mesons $\lambda_V$ are determined from the radiative decay widths
$\Gamma_{V\rightarrow\eta\gamma}$ via
\begin{equation}
\Gamma_{V\rightarrow\eta\gamma} =
\frac{\alpha(m_V^2-m_\eta^2)^3}{24\,m_V^3\,m_\eta^2}\,\lambda_V^2 \,,
\end{equation}
and the electromagnetic form factor $F_V^{em}(Q^2)$ is assumed to have the
usual dipole behavior. The off-shell behavior of the hadronic couplings
$g_v$ (vector coupling) and $g_t$ (tensor coupling) in Eq.~(\ref{eq:LVNN})
is described by a dipole form factor
\begin{equation}
 g_{v,\,t} = \tilde{g}_{v,\,t}\, \frac{\left(\Lambda_V^2 - m_V^2\right)^2}
 {\left(\Lambda_V^2 + \bm{k}_V^2 \right)^2} \,.
\end{equation}
In general the values for the strong coupling constants $\tilde{g}_v$  and
$\tilde{g}_t$ are not well determined. In various
analyses~\cite{Drechsel:1998hk,Davidson:1991xz,Dumbrajs:1983jd} they vary
in the ranges of $1.8 \leq \tilde{g}^\rho_v \leq 3.2$, $8\leq
\tilde{g}^\omega_v \leq 20$, $4.3 \leq \tilde{g}^\rho_t/\tilde{g}^\rho_v
\leq 6.6$, and $-1 \leq \tilde{g}^\omega_t/\tilde{g}^\omega_v \leq 0$. In
the present work we take them as free parameters to be varied within these
ranges. The fitted values of these parameters are given in
Table~\ref{tbl:vecmes}.

\begin{table}
\caption{Parameters for the vector mesons.} \label{tbl:vecmes}%
\smallskip
\begin{tabular}{c c c c c c}
\hline
 $V$ & $m_V\,[{\mathrm MeV}]$ & $\tilde{g}_v$ & $\tilde{g}_t/\tilde{g}_v$ &
 $\Lambda_V\,[{\mathrm GeV}]$ & $\lambda_V$ \\
\hline
 $\rho$   & $768.5$ & $ 2.4$ & $6.1$ & $1.3$ & $0.810$ \\
 $\omega$ & $782.6$ & $16  $ & $0  $ & $1.3$ & $0.291$ \\
\hline
\end{tabular}
\end{table}

\subsection{Resonance contribution} \label{sec:Reson}
In addition to the dominant $S_{11}(1535)$, we also consider $N^*$
contributions from $D_{13}(1520)$, $S_{11}(1650)$, $D_{15}(1675)$,
$F_{15}(1680)$, $D_{13}(1700)$, $P_{11}(1710)$, and $P_{13}(1720)$. For the
relevant multipoles $\mathcal{M}_{\ell\pm}$ ($=E_{\ell\pm},\,
M_{\ell\pm},\, S_{\ell\pm}$) of the resonance contributions, we assume a
Breit-Wigner energy dependence of the form
\begin{equation} \label{eq:BWres}
 \mathcal{M}_{\ell\pm}(W,Q^2) = \tilde{\mathcal{M}}_{\ell\pm}(Q^2)\,
 \frac{W_R \Gamma_\mathrm{tot}(W)}{W_R^2-W^2-i W_R \Gamma_\mathrm{tot}(W)}\,
 f_{\eta N}(W)\, C_{\eta N} \,,
\end{equation}
where $f_{\eta N}(W)$ is the usual Breit-Wigner factor describing the $\eta
N$ decay of the $N^*$ resonance with total width $\Gamma_\mathrm{tot}$,
partial width $\Gamma_{\eta N}$ and spin $J$,
\begin{equation} \label{eq:fetaN}
 f_{\eta N}(W) = \zeta_{\eta N} \left[ \frac{1}{(2J+1)\pi}\,
 \frac{k_W}{|\,\bm{q}\,|}\, \frac{m_N}{W_R}\,
 \frac{\Gamma_{\eta N}}{\Gamma_\mathrm{tot}^2} \right]^{1/2},
 \quad k_W = \frac{W^2-m_N^2}{2W}\,,
\end{equation}
where $\zeta_{\eta N} = \pm 1$ describes the relative sign between $N^*
\rightarrow \eta N$ and $N^* \rightarrow \pi N$ couplings. The isospin
factor $C_{\eta N}$ is $-1$, and $\tilde{E}_{\ell\pm}$,
$\tilde{M}_{\ell\pm}$ and $\tilde{S}_{\ell\pm}$ are related to the photon
excitation helicity amplitudes by
\begin{eqnarray}
 A^{\ell+}_{1/2} &=& -\frac{1}{2}\, \bigl[ (\ell+2)\tilde{E}_{\ell+}
 + \ell\,\tilde{M}_{\ell+} \bigr] ,
 \nonumber \\ [.5ex]%
 A^{\ell+}_{3/2} &=& \frac{1}{2}\, \sqrt{\ell(\ell+2)}\,
 \bigl(\tilde{E}_{\ell+} - \tilde{M}_{\ell+} \bigr) ,
 \nonumber \\ [.5ex]%
 A^{(\ell+1)-}_{1/2} &=& -\frac{1}{2}\, \bigl[ \ell\,\tilde{E}_{(\ell+1)-}
 - (\ell+2)\tilde{M}_{(\ell+1)-} \bigr] ,
 \nonumber \\ [.5ex]%
 A^{(\ell+1)-}_{3/2} &=& -\frac{1}{2}\, \sqrt{\ell(\ell+2)}\,
 \bigl( \tilde{E}_{(\ell+1)-} + \tilde{M}_{(\ell+1)-} \bigr) ,
 \nonumber \\ [0.5ex]%
 S^{\ell+}_{1/2} &=& -\frac{1}{\sqrt{2}}\, (\ell+1)\, \tilde{S}_{\ell+} \,,
 \nonumber \\ [0.5ex]%
 S^{(\ell+1)-}_{1/2} &=&
                 -\frac{1}{\sqrt{2}}\, (\ell+1)\, \tilde{S}_{(\ell+1)-} \,.
\end{eqnarray}
The scalar multipole amplitudes $S_{\ell\pm}$ appear only in
electroproduction, and are related to the longitudinal ones by $\omega
S_{\ell\pm} = |\bm{k}| L_{\ell\pm}$.

In accordance with Ref.~\cite{Walker:1969xu}, the energy dependence of the
partial width $\Gamma_{\eta N}$ is given by
\begin{equation} \label{eq:GametaN}
 \Gamma_{\eta N}(W) = \beta_{\eta N }\,\Gamma_R
 \left(\frac{|\;\bm{q}\;|}{|\bm{q}_R|}\right)^{2\ell+1}
 \left(\frac{X^2+\bm{q}_R^2}{X^2+\bm{q}^2}\right)^\ell \frac{W_R}{W}\,,
\end{equation}
where $X$ is a damping parameter, assumed to be $500$ MeV for all
resonances. $\Gamma_R$ and $q_R$ are the total width and the $\eta$ c.m.
momentum at the resonance peak ($W=W_R$) respectively, and $\beta_{\eta N}$
is the $\eta N$ decay branching ratio.

The total width $\Gamma_\mathrm{tot}$ in Eqs.~(\ref{eq:BWres}) and
(\ref{eq:fetaN}) is the sum of $\Gamma_{\eta N}$, the single-pion decay
width $\Gamma_{\pi N}$, and the rest, for which we assume dominance of the
two-pion decay channels,
\begin{equation}
 \Gamma_\mathrm{tot}(W) =
 \Gamma_{\eta N}(W) + \Gamma_{\pi N}(W) + \Gamma_{\pi\pi N}(W)\,.
\end{equation}
The width $\Gamma_{\pi N}$ has a similar energy dependence as
$\Gamma_{\eta N}$, and $\Gamma_{\pi\pi N}$ is parameterized in an energy
dependent form,
\begin{eqnarray}
 \Gamma_{\pi N}(W) &=& \beta_{\pi N }\,\Gamma_R
 \left(\frac{|\bm{q_\pi}|}{|\bm{q}_R|}\right)^{2\ell+1}
 \left(\frac{X^2+\bm{q}_R^2}{X^2+\bm{q_\pi}^2}\right)^\ell \frac{W_R}{W} \,,
 \\
 \Gamma_{\pi\pi N}(W) &=& (1-\beta_{\pi N}-\beta_{\eta N})\, \Gamma_R
 \left(\frac{q_{2\pi}}{q_0}\right)^{2\ell+4}
 \left(\frac{X^2+q_0^2}{X^2+q_{2\pi}^2}\right)^{\ell+2} ,
\end{eqnarray}
where $q_{2\pi}$ is the momentum of the compound (2$\pi$) system with mass
$2m_{\pi}$ and $q_0=q_{2\pi}$ at $W=W_R$. The definition of $\Gamma_{\pi\pi
N}$ has been chosen to account for the correct energy behavior of the phase
space near the three-body threshold.

\begin{table}
\caption{Parameters of nucleon resonances studied in our isobar model. The
masses and widths are given in MeV, $\zeta_{\eta N}$ give the relative sign
between $N^* \rightarrow \eta N$ and $N^* \rightarrow \pi N$ couplings, and
$\beta_{\eta N,\,\pi N,\,\pi\pi N}$ are the branching ratios for the
respective decay channels. In the first row for each resonance, we list the
average values or ranges given by the Particle Data Group
(PDG)~\cite{PDG:2000}. The numbers in the second rows are the values used
in our model, the underlined ones being taken as free parameters determined
from data fitting. If no value is shown in the second rows, then the PDG
value is adopted.}
\label{tbl:ResPar1}%
\smallskip
\renewcommand{\arraystretch}{1.2}
\begin{tabular}{ccccccc}  \hline
 $N^*$  &  Mass  &  Width  &  $\zeta_{\eta N}$
 & $\beta_{\eta N}$  &  $\beta_{\pi N}$  &  $\beta_{\pi\pi N}$ \\
 \hline
 $D_{13}(1520)$ & 1520  & 120 & $+1$ & $0.08\pm0.01\%$      & $50-60\%$ & $40-50\%$\\
        &     &     &    & \underline{$0.06\%$} & $60\%$ & $40\%$\\ [1ex]
 $S_{11}(1535)$ & 1520-1555 & 100-250& $+1$ & $30-55\%$& $35-55\%$& $1-10\%$ \\
        & \underline{1541}& \underline{191}&    & $50\%$& $40\%$& $10\%$\\ [1ex]
 $S_{11}(1650)$ & 1640-1680 & 145-190 & $-1$ & $3-10\%$ & $55-90\%$ & $10-20\%$ \\
        & \underline{1638}  & \underline{114} &
        & \underline{$7.9\%$} & $77\%$ & $15\%$ \\ [1ex]
 $D_{15}(1675)$ & 1670-1685 & 150 & $-1$ & $0.1\pm0.1\%$& $40-50\%$& $50-60\%$ \\
        & \underline{1665} &   &   & \underline{$17\%$}  & $40\%$& $43\%$ \\ [1ex]
 $F_{15}(1680)$ & 1675-1690 & 130 & $+1$ & $0.15\pm0.3\%$& $60-70\%$& $30-40\%$ \\
        & \underline{1681} &   &   & \underline{$0.06\%$}& $60\%$& $40\%$ \\ [1ex]
 $D_{13}(1700)$ & 1700 & 100& $-1$ & $10\pm6\%$ & $5-15\%$ & $85-95\%$\\
        &    &    &    & \underline{$0.3\%$} & $15\%$ & $85\%$ \\ [1ex]
 $P_{11}(1710)$ & 1680-1740 & 100 & $+1$ & $16\pm10\%$ & $10-20\%$ & $40-90\%$\\
        & \underline{1721}&   &   & \underline{$26\%$}& $14\%$ & $60\%$ \\ [1ex]
 $P_{13}(1720)$ & 1720      & 150 & $+1$ & $0.2\pm1\%$ & $10-20\%$ & $>70\%$ \\
        &    &    &    & \underline{$3.0\%$} & $15\%$ & $82\%$ \\
 \hline
\end{tabular}
\end{table}

\begin{table}
\caption{Photoexcitation helicity amplitudes $A^p_{1/2,\,3/2}$ (in
$10^{-3}\;\mbox{GeV}^{-1/2}$) of nucleon resonances studied in our isobar
model. The PDG values are given in the first row for each resonance, and
our fitting results are underlined in the second row. The $A^p_{1/2}$ for
$S_{11}(1650)$ is obtained through the relation of Eq.~(\ref{eq:2S11}). The
quantities $\xi_{1/2,\,3/2}$ (in $10^{-1}\;\mbox{GeV}^{-1}$) are defined in
Eq.~(\ref{eq:xi}).}
\label{tbl:ResPar2}%
\smallskip
\renewcommand{\arraystretch}{1.2}
\begin{tabular}{cccrr}  \hline
 $N^*$ & $A^p_{1/2}$ & $A^p_{3/2}$ & $\xi_{1/2}$~ & $\xi_{3/2}$~ \\
 \hline
 $D_{13}(1520)$& $- 24\pm 9$      & $+166\pm 5$      &       &      \\
               &\underline{$- 52$}&                  &-0.049 & 0.155\\ [1ex]
 $S_{11}(1535)$& $+ 90\pm30$      & ---              &       &      \\
               &\underline{$+118$}& ---    & 2.34\phantom{0} & ---~~\\ [1ex]
 $S_{11}(1650)$& $+ 53\pm16$      & ---              &       &      \\
               & $+ 68     $      & ---              & 0.554 & ---~~\\ [1ex]
 $D_{15}(1675)$& $+ 19\pm 8$      &  $+ 15\pm 9$     &       &      \\
               &\underline{$+ 18$}&\underline{$+ 24$}& 0.178 & 0.242\\ [1ex]
 $F_{15}(1680)$& $- 15\pm 6$      & $+133\pm12$      &       &      \\
               &\underline{$- 21$}&\underline{$+125$}&-0.013 & 0.078\\ [1ex]
 $D_{13}(1700)$& $- 18\pm13$      & $-  2\pm24$      &       &      \\
               &                  &                  &-0.028 &-0.003\\ [1ex]
 $P_{11}(1710)$& $+  9\pm22$      & ---              &       &      \\
               &\underline{$+ 23$}& ---              & 0.329 & ---~~\\ [1ex]
 $P_{13}(1720)$& $+ 18\pm30$      &  $- 19\pm20$     &       &      \\
               &                  &                  & 0.071 &-0.075\\
 \hline
\end{tabular}
\end{table}

Differently from the treatment of resonance contributions in MAID we do not
use the form factor $f_{\gamma N}(W)$ for the $\gamma NN$ vertex in the
Eq.~(18) of Ref.~\cite{Drechsel:1998hk}. The resonance parameters in this
isobar model are given in Tables~\ref{tbl:ResPar1} and \ref{tbl:ResPar2}.

\subsection{Electroproduction} \label{sec:Elec}
For the $Q^2$ dependence of the $S_{11}$(1535) resonance we assume the form
\begin{equation} \label{eq:S11Q2}
  A_{1/2}^p(Q^2) = A_{1/2}^p(0)\,\frac{1 + s_n\,Q^2}{1 + s_d\,Q^2}\,
                   F_D(Q^2) \,,
\end{equation}
where $s_n$ and $s_d$ are taken as parameters to be determined, and
$F_D(Q^2)$ is the standard nucleon dipole form factor. As in the case of
photoproduction, we follow the assumption in the single quark transition
model~\cite{Burkert:1992} for the second $S_{11}$(1650) resonance,
\begin{equation} \label{eq:2S11}
  A_{1/2}^{S_{11}(1650)}(Q^2) = A_{1/2}^{S_{11}(1535)}(Q^2)\,\tan30^\circ\,.
\end{equation}

For $D_{13}$(1520) and $F_{15}$(1680), we follow the MAID except we modify
the $Q^2 \leq 1\;(\mathrm{GeV/c})^2$ region to comply our photoproduction
($Q=0$) fitted values.

For the rest of the resonances involved in this study, we take the
following form for their multipoles
\begin{equation} \label{eq:Q2}
  \tilde{\mathcal{M}}_{\ell\pm}(Q^2) = \tilde{\mathcal{M}}_{\ell\pm}(0)
  \frac{|\,\bm{k}\,|}{k_W} F_D(Q^2)\,.
\end{equation}
Therefore, we fit the electroproduction data with only two new parameters,
$s_n$ and $s_d$. The rest of the parameters are fixed from the
photoproduction fit.

\section{Results and discussion} \label{sec:Result}

\subsection{Photoproduction results} \label{sec:PhoRes}
Applying our isobar model, we have fitted recent photoproduction data
including total and differential cross sections from TAPS
(MAMI/Mainz)~\cite{Krusche:1995nv} and GRAAL~\cite{Renard:2000iv}, as well
as the polarized beam asymmetry from GRAAL~\cite{Ajaka:1998zi}. Although
the polarized target asymmetry has been measured at ELSA
(Bonn)~\cite{Bock:1998rk}, we did not include it in our fitting for the
reason which will be discussed later. Instead, we compare our prediction
with these Bonn data. In Tables~\ref{tbl:ResPar1} and \ref{tbl:ResPar2} we
show our fit results of resonance parameters (second row) and compare with
the PDG values~\cite{PDG:2000} (first row). To reduce the number of
parameters in this analysis, the PDG value for a resonance parameter is
adopted when the parameter is not found to be sensitive for the fit to the
current experimental data.

\subsection{Cross sections} \label{sec:XSec}
\begin{figure}
  \centering
  \includegraphics[width=9cm]{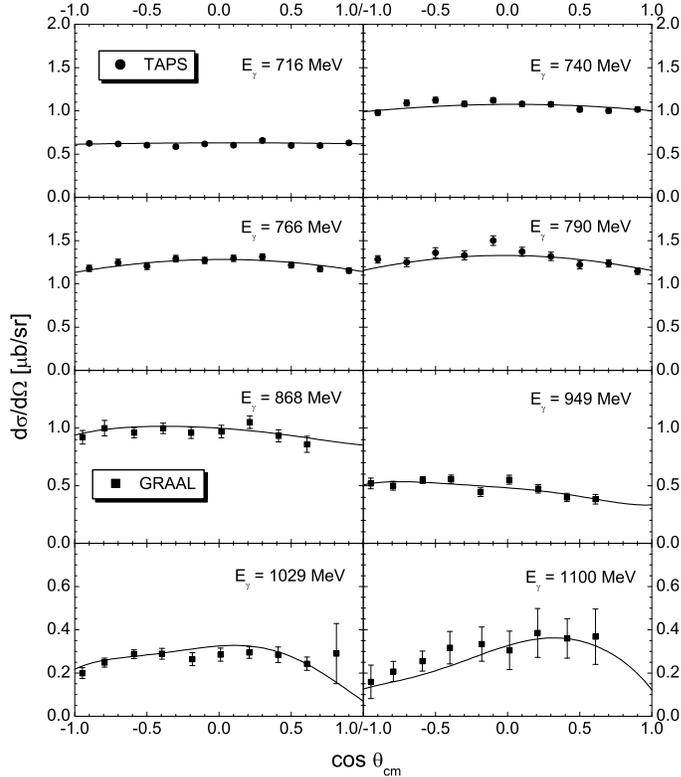}
  \caption{Differential cross section for $\gamma p\rightarrow \eta p$.
  The data are from TAPS~\cite{Krusche:1995nv} and
  GRAAL~\cite{Renard:2000iv}.}
  \label{fig:dcs}
\end{figure}

The TAPS data~\cite{Krusche:1995nv} include differential cross sections as
well as total cross sections from threshold ($E_\gamma^\mathrm{lab}$ = 707
MeV) up to $E_\gamma^\mathrm{lab}$ = 790 MeV, which is nearly the peak of
$S_{11}(1535)$. The GRAAL data~\cite{Renard:2000iv} contain differential
cross sections measured from threshold to $E_\gamma^\mathrm{lab}$ = 1100
MeV. These data cover a wider energy region, but do not provide total cross
sections independently, instead these are obtained by integration of the
differential cross sections. These data sets are in good agreement with
each other in the overlapping energy region. In Fig.~\ref{fig:dcs}, our
results for differential cross sections are in very good agreement with the
data from TAPS and GRAAL. In the low energy region the differential cross
section is flat, indicating $s$-wave dominance. As the energy increases,
higher partial waves start to contribute. Note that our results at
$E_\gamma^\mathrm{lab}> 1\ \mbox{GeV}$ show a dropping behavior at forward
angles, which is not seen in the GRAAL data, but still within their error
bars.

\begin{figure}
  \centering
  \includegraphics[width=9cm]{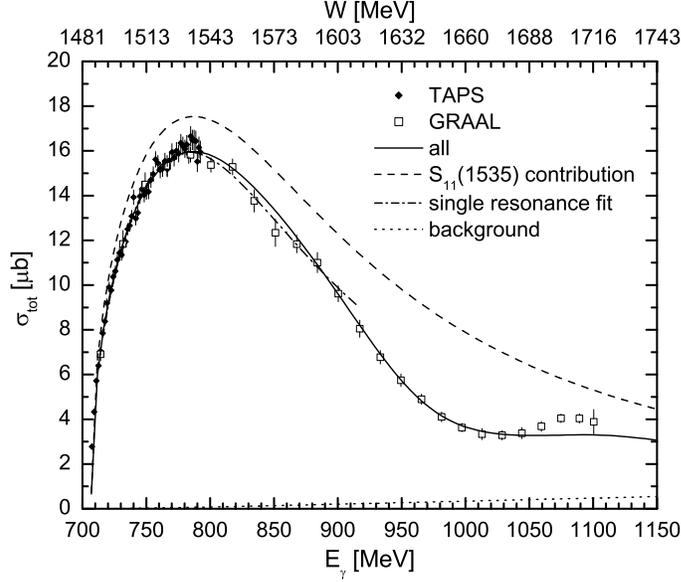}
  \caption{Total cross section for $\gamma p\rightarrow \eta p$.
  The data are from TAPS~\cite{Krusche:1995nv} and
  GRAAL~\cite{Renard:2000iv}.}
  \label{fig:tcs}
\end{figure}

\begin{figure}
  \centering
  \includegraphics[width=9cm]{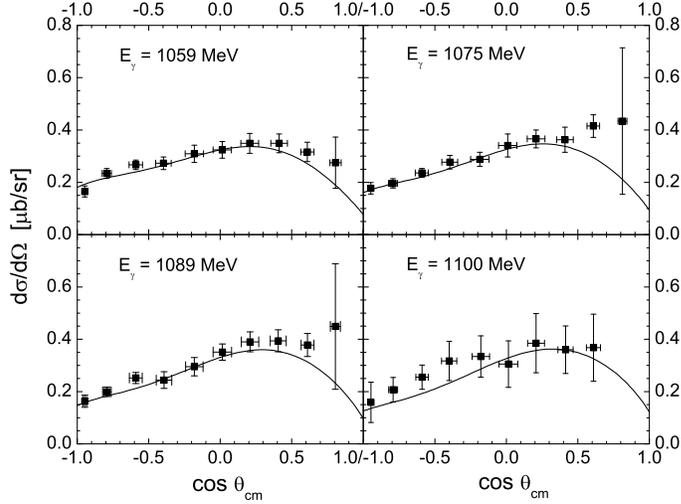}
  \caption{Differential cross section for $\gamma p\rightarrow \eta p$ at
  energies between $E_\gamma^\mathrm{lab}$ = 1050 - 1100 MeV.
  The data are from GRAAL~\cite{Renard:2000iv}.}
  \label{fig:tcshe}
\end{figure}

Our results for the total cross section is shown in Fig.~\ref{fig:tcs}, and
compared with the TAPS and GRAAL data. Again, these are in good agreement
except for the bump observed in the GRAAL data in the region
$E_\gamma^\mathrm{lab}$ = 1050 - 1100 MeV that can not be reproduced in our
model. However, note that the total cross section of the GRAAL data is
obtained from integrating the differential cross sections, by use of a
polynomial fit in $\cos\theta$ for extrapolation to the uncovered region.
Therefore, the proper way to figure out the discrepancy is to compare the
differential cross sections directly. In Fig.~\ref{fig:tcshe}, we plot the
differential cross sections at four different energies between
$E_\gamma^\mathrm{lab}$ = 1050 - 1100 MeV where the bump occurs. It is seen
from these differential cross sections that there are no obvious
differences, except for the forward angles, where the error bars are quite
big. Therefore, we conclude that the discrepancy is due to the
extrapolation of the GRAAL data to the forward angles and not heavily
supported by the data themselves.

\begin{table}
\caption{Parameters of the $S_{11}(1535)$ resonance. The results of the
single resonance fit are obtained by performing the fit to low energy data
using only one $S_{11}$ resonance. The second row is the our full result
including two $S_{11}$ resonances.}
\label{tbl:SRFit}%
\smallskip
\renewcommand{\arraystretch}{1.4}
\begin{tabular}{cccccc}
\hline
 Fit & Mass  & Width & $A^p_{1/2}$ & $\beta_{\eta N}^{S_{11}(1535)}$ \\
     & [MeV] & [MeV] & [$10^{-3}\ \mbox{GeV}^{-1/2}$] &   \\
 \hline
 Single $S_{11}$ resonance fit & 1536 & 159 & 103 & $50\%$  \\ [0.3ex]
 Double $S_{11}$ resonance fit & 1541 & 191 & 118 & $50\%$  \\
 \hline
\end{tabular}
\end{table}

Fig.~\ref{fig:tcs} shows that the background contribution is very small,
and the total cross section is dominated by the $S_{11}(1535)$ at low
energy. However, the contribution from the second resonance,
$S_{11}(1650)$, can not be neglected. Even though a single $S_{11}$
resonance can fit the low energy data nicely up to $E_\gamma^\mathrm{lab}$
= 910 MeV (the dash-dotted curve in Fig.~\ref{fig:tcs}), it can by no means
describe the higher energy region. Moreover, the single resonance fit
yields incorrect resonance parameters, as shown in Table~\ref{tbl:SRFit}.
In fact, the decay width and photon coupling obtained in the single
$S_{11}$ resonance fit are significantly smaller than the full results when
both $S_{11}$ resonances are properly included.

\subsection{Polarization observables} \label{sec:SpinObs}

One special feature in the polarization measurements of $\eta$
photoproduction is that one can access small contributions from a
particular resonance through the interference of the dominant $E_{0+}$
multipole with smaller multipoles. For example, if we assume the $s$-wave
dominance, the polarized beam asymmetry can be expressed as
\begin{equation}\label{eq:Sigma}
  \Sigma = 3\,\sin^2\theta\;
  \mathrm{Re}[E_{0+}^*(E_{2-}+M_{2-})]\, / \,|E_{0+}|^2\,,
\end{equation}
which only depends on the interference of $E_{0+}$ and $(E_{2-}+M_{2-})$
multipoles. Since the background is very small in this reaction, the main
source producing $E_{2-}$ and $M_{2-}$ at low energy is the $D_{13}$(1520)
resonance. This is the reason why the photon asymmetry is so sensitive to
the $D_{13}$(1520) at low energies, and why even the tiny branching ratio
can be determined.

\begin{figure}
  \centering
  \includegraphics[width=9cm]{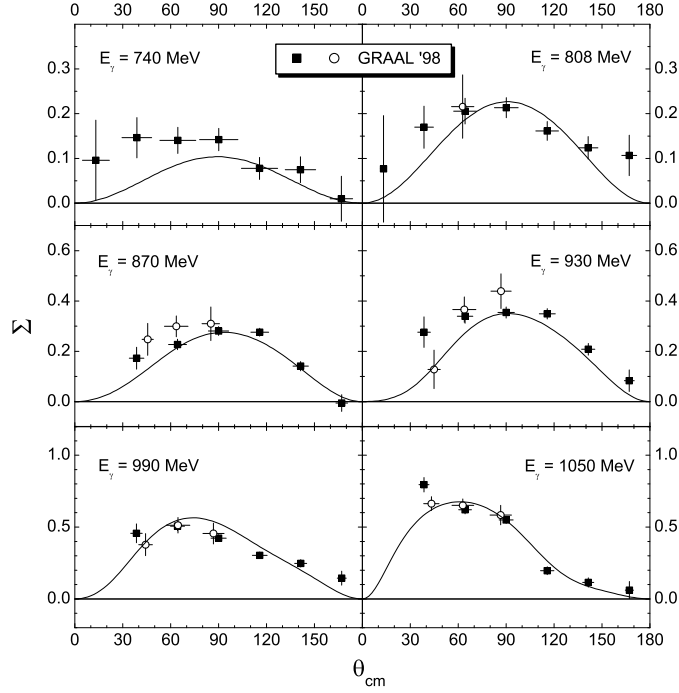}
  \caption{Beam asymmetry for $\gamma p\rightarrow \eta p$.
  The data are from GRAAL~\cite{Ajaka:1998zi}.}
  \label{fig:beam}
\end{figure}

The available beam asymmetry data were measured at
GRAAL~\cite{Ajaka:1998zi} from threshold to $E_\gamma^\mathrm{lab}$ = 1.1
GeV. Higher energy data up to $E_\gamma^\mathrm{lab}$ = 1.5~{GeV} are being
analyzed and will be available soon~\cite{D'Angelo:2001}. In
Fig.~\ref{fig:beam}, we compare our results with these data. An overall
good agreement has been achieved. At low energies, we observe that the beam
asymmetry has a clear $\sin^2\theta$ dependence, which behaves according to
Eq.~(\ref{eq:Sigma}) as a result of interference between $s$- and
$d$-waves. At the lowest energy (740~MeV) the data show an asymmetric
forward-backward (f-b) shape which can not be described with our model. In
fact, due to the large suppression of high partial waves this would be true
for any model.  From these low energy data, a branching ratio of
$\beta_{\eta N} = (0.06 \pm 0.003)\%$ can be determined for the
$D_{13}$(1520). When energies get higher than $E_\gamma^\mathrm{lab}$ =
930~{MeV}, the data develop a f-b asymmetry behavior, which becomes
especially evident at $E_\gamma^\mathrm{lab}$ = 1050~{MeV}. This f-b
asymmetry in $\Sigma$ is very sensitive to the $F_{15}$(1680) as discussed
by Tiator {\it et al.}~\cite{Tiator:1999gr}, which is the reason why such a
small branching ratio $(0.06 \pm 0.02)\%$ can be extracted for this
resonance. The same f-b asymmetry behavior is also responsible for the
large branching ratio of the $D_{15}$(1675) (17\%) in our fit.

\begin{figure}
  \centering
  \includegraphics[width=9cm]{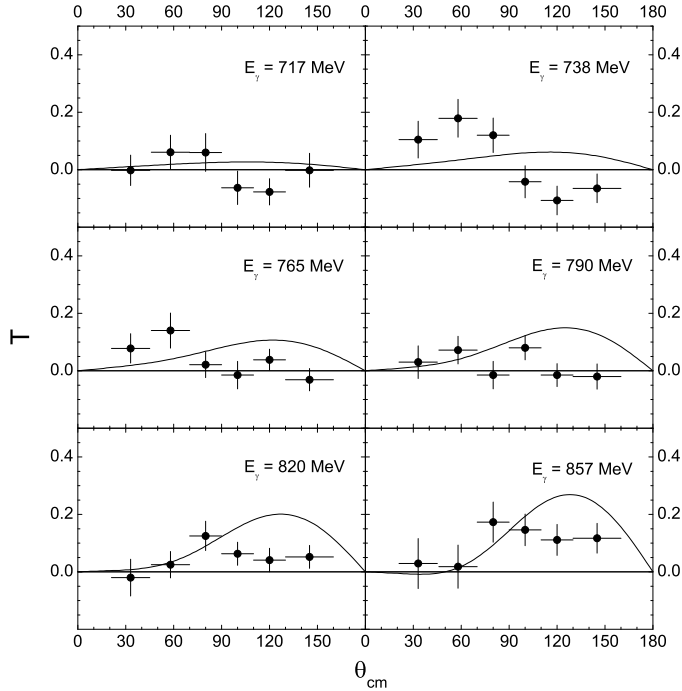}
  \caption{Target asymmetry for $\gamma p\rightarrow \eta p$.
  The data are from Bonn~\cite{Bock:1998rk}.}
  \label{fig:target}
\end{figure}

The target asymmetry for $\gamma p\rightarrow \eta p$ has been measured at
Bonn~\cite{Bock:1998rk} from threshold to $E_\gamma^\mathrm{lab}$ = 1150
MeV. However, these data are \emph{not} included in our current fitting
because so far all theoretical efforts fail to reproduce them. Instead, we
plot the prediction from our model in Fig.~\ref{fig:target} and compare
with the Bonn data. At low energies, a nodal structure with sign changes at
around $90^\circ$ is observed from these Bonn data, but is not present in
our result. At higher energies, our results for the target asymmetry show
an enhancement at forward angles, mainly due to the contribution from the
$D_{15}(1675)$. The discrepancy at low energies also occurs in other
models~\cite{Feuster:1998cj,Mukhopadhyay:1998hn,Li:1998ni,Saghai:2001yd},
and has been discussed in detail in Ref.~\cite{Tiator:1999gr}, where it was
pointed out that an unexpectedly large relative phase difference between
$s$- and $d$-waves is required to explain these Bonn data. Such target
asymmetry results, especially the nodal structure at low energies, need to
be confirmed by further experiments, and such experiments have been
proposed at GRAAL and MAMI. These target asymmetry results should reveal
further properties of less-established resonances, as has been already
demonstrated by the beam asymmetry results.

\begin{figure}
  \centering
  \includegraphics[width=9cm]{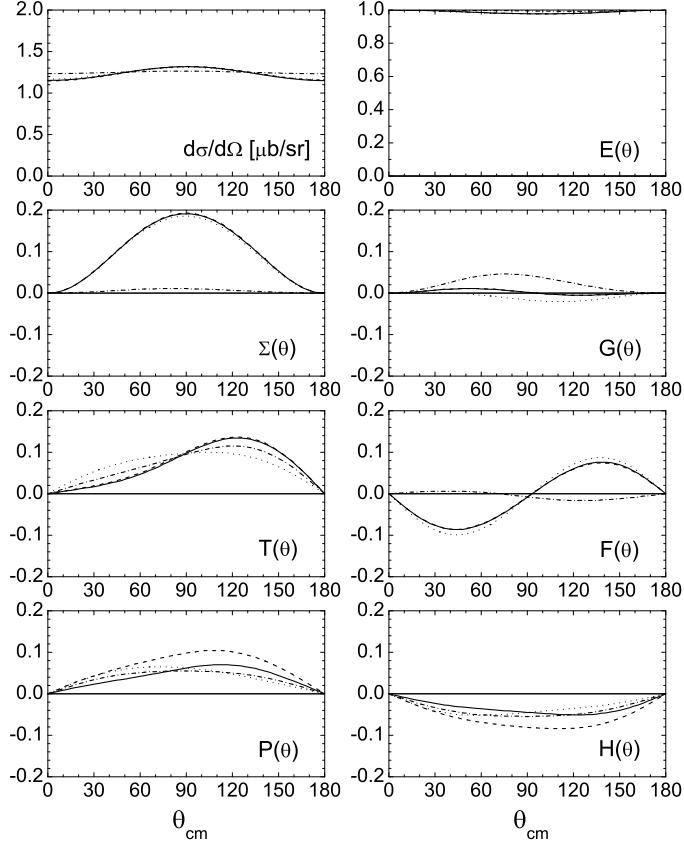}
  \caption{Influence of resonances on the differential cross
  section,
  the single-polarization observables $\Sigma$, $T$ and $P$, and the
  double-polarization (beam-target) observables $E$, $F$, $G$, $H$
  at a center-of-mass energy $W=1530$ MeV. The solid line is the
  full isobar model result. The dashed line is obtained without the
  contributions from the $P_{11}(1710)$, the dotted line without the
  $D_{15}(1675)$, and the dash-dotted line without both $D_{13}(1520)$ and
  $D_{13}(1700)$.}
  \label{fig:pol1530}
\end{figure}

\begin{figure}
  \centering
  \includegraphics[width=9cm]{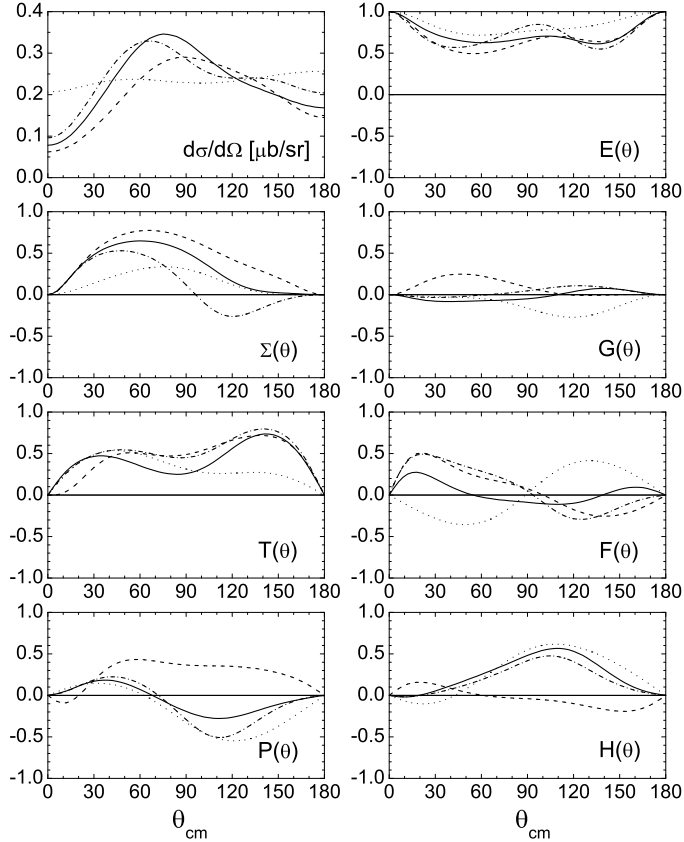}
  \caption{Influence of resonances on the differential cross section and
  polarization observables at a center-of-mass energy $W=1700$ MeV.
  Notation of the curves as in Fig.~\ref{fig:pol1530}.}
  \label{fig:pol1700}
\end{figure}

\begin{figure}
  \centering
  \includegraphics[width=9.5cm]{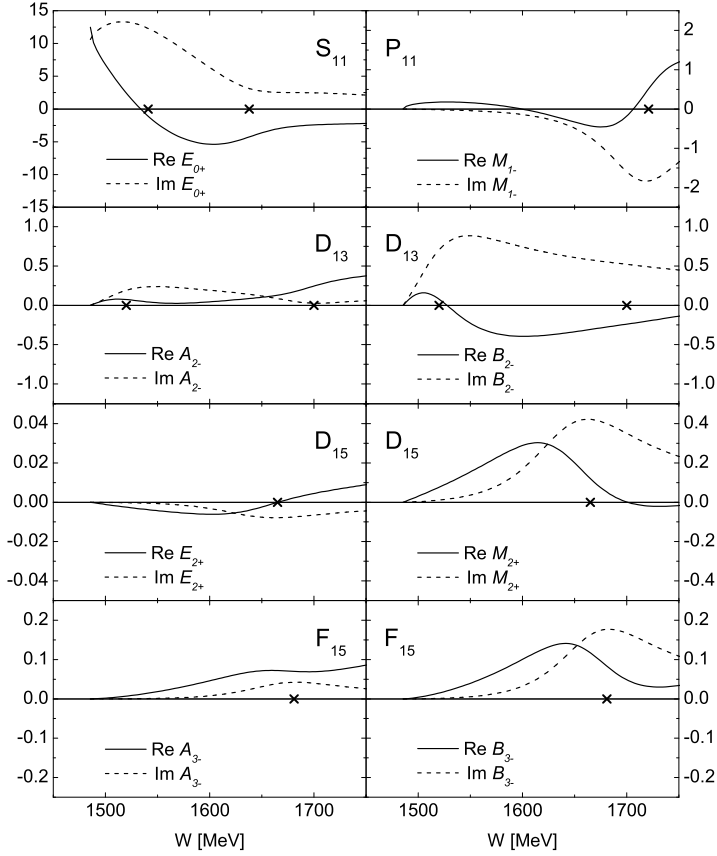}
  \caption{Real and imaginary parts of $E_{0+}$, $M_{1-}$, $E_{2+}$ and
  $M_{2+}$ multipoles and $A_{2-}=(3M_{2-}-E_{2-})/2$,
  $B_{2-}=M_{2-}+E_{2-}$, $A_{3-}=2M_{3-}-E_{3-}$ and $B_{3-}=M_{3-}+E_{3-}$
  helicity amplitudes (in $10^{-3}/m_{\pi^+}$) for the reaction
  $\gamma p\rightarrow \eta p$. The nucleon resonances $S_{11}(1535)$,
  $S_{11}(1650)$, $P_{11}(1710)$, $D_{13}(1520)$, $D_{13}(1700)$,
  $D_{15}(1675)$ and $F_{15}(1680)$ are indicated by crosses at their mass
  positions in the corresponding partial channels.}
  \label{fig:multipol}
\end{figure}

In Figs.~\ref{fig:pol1530} and \ref{fig:pol1700} we show two different
energy sets of 8 out of 16 possible polarization observables for
photoproduction which are currently under consideration. All of them can be
reached with a polarized target and linearly or circularly polarized
photons. Even the recoil polarization $P$ can be obtained with a polarized
target using double polarization. At $W=1530\;\mbox{MeV}$, in the center of
the $S_{11}(1535)$ and $D_{13}(1520)$ resonances, the behavior is very
strongly dominated by the large $s$-wave $E_{0+}$. As can be seen in
Fig.~\ref{fig:multipol},
in this region the $s$-wave dominates by even more than a magnitude and for
the $p$- and $d$-waves the only sizeable contribution arises from the
$B_{2-}$ multipole, the helicity $3/2$ component of the $D_{13}(1520)$
resonance excitation amplitude. Consequently, the differential cross
section is practically constant and the helicity asymmetry $E(\theta)$ is
unity. In this energy domain $E(\theta)$ can well be used for calibration
purposes of circularly polarized photons as this result is highly model
independent. For a pure $s$-wave all other polarization observables in
Fig.~\ref{fig:pol1530} would completely vanish and the values obtained are
interferences between the $s$-wave and other resonance and background
contributions, where all of them are relatively small. The photon asymmetry
$\Sigma$ and the double polarization observable $F$ are strongly dominated
by the $s$-$d$ interference $\mbox{Re}[E_{0+}^*B_{2-}]$ while all other
resonance contributions are negligible. A special case is the target
polarization $T$ which is proportional to $\mbox{Im}[E_{0+}^*B_{2-}]$.
However, as the two resonances are so close together (within 10-20 MeV),
the imaginary part disappears in our isobar model. As discussed before,
this observable should be of very high priority for a next generation of
eta photoproduction experiments. The higher resonances $D_{15}(1675)$,
$D_{13}(1700)$ and $P_{11}(1710)$ do not significantly contribute at this
low energy, giving the possibility to study the background amplitudes in
the target and recoil polarizations $T$ and $P$. The situation changes
dramatically in Fig.~\ref{fig:pol1700} at $W=1700\;\mbox{MeV}$. The
$s$-wave no longer dominates so strongly (see Fig.~\ref{fig:multipol}) and
higher resonances play bigger roles. The cross section and $E(\theta)$
exhibit structures, the $D_{15}(1675)$ is mostly responsible for the
angular shape of the cross section. It also plays a big role in the
$\Sigma$, $T$ and $F$ observable. As in the case of pion production, it is
generally very difficult to find enhanced sensitivity to $P_{11}$
resonances, like the Roper. In most observables the influence of the
$M_{1-}$ multipole is very small. Here in Fig.~\ref{fig:pol1700} the recoil
polarization and the double polarization observable $H$ give access to
study the $P_{11}(1710)$ resonance.

In Fig.~\ref{fig:multipol} the most important partial waves are shown as
real and imaginary parts of electric and magnetic multipoles or as helicity
1/2 and 3/2 amplitudes. The $D_{13}(1520)$ and $F_{15}(1680)$ resonances
are the two most prominent states with large $A_{3/2}$ photon couplings at
$Q^2=0$. This can be very well seen in the $B_{2-}$ and $B_{3-}$
amplitudes. On the other hand, the $D_{15}(1675)$ has similar $A_{1/2}$ and
$A_{3/2}$ couplings (see Table~\ref{tbl:ResPar2}) but is very strongly
dominated by magnetic transitions. The fact that in general the real parts
of the amplitudes do not vanish at the resonance positions as expected from
the Breit-Wigner forms shows the influence of the background contributions.

\subsection{Electroproduction results} \label{sec:ElecRes}
There are two recent $\eta$ electroproduction data sets from Jefferson Lab:
Armstrong {\it et al.}~\cite{Armstrong:1998wg} measured the $e p
\rightarrow e' p \eta$ process at high momentum transfer ($Q^2 =
2.4,\,3.6\; (\mbox{GeV/c})^2$) and $W$ around the $S_{11}(1535)$ region,
while the CLAS collaboration~\cite{Thompson:2000by} measured at various
$Q^2$ ($=0.25-1.5\; (\mbox{GeV/c})^2$) over a wider energy region
($W=1.5-1.86\;\mbox{GeV}$). When fitting these electroproduction data, we
fix all the parameters determined from the photoproduction data except for
the $Q^2$ dependence of the helicity amplitudes $A^p_{1/2,\,3/2}(Q^2)$. The
$Q^2$ dependence of the $S_{11}(1535)$ is described by the form of
Eq.~(\ref{eq:S11Q2}), and the parameters $s_n$ and $s_d$ are determined
from electroproduction data. Fitting all the JLab data, we obtain the
results: $s_n=2.394\;(\mbox{GeV/c})^{-2}$ and
$s_d=0.085\;(\mbox{GeV/c})^{-2}$.

\begin{figure}
  \centering
  \includegraphics[width=8cm]{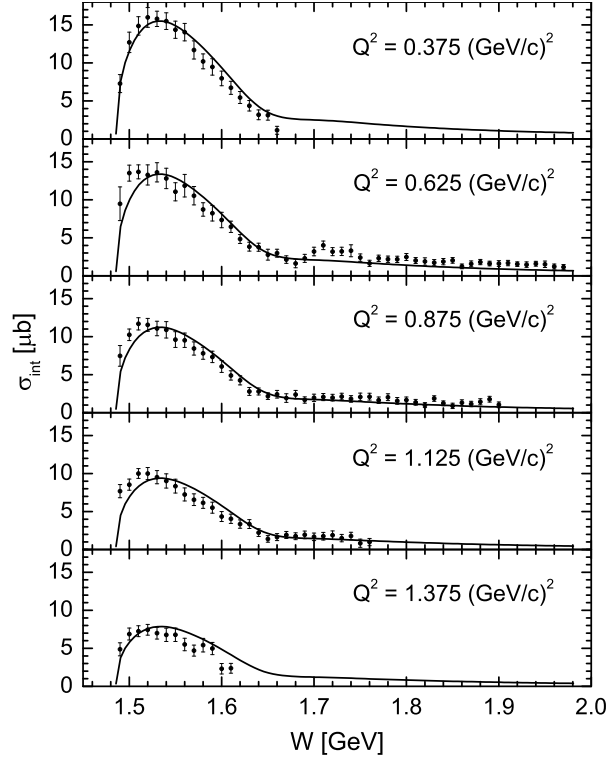}
  \caption{Total cross section for $p\,(e,e'p)\,\eta$ at various
  $Q^2$ compared to the CLAS integrated cross section
  data~\cite{Thompson:2000by}.}
  \label{fig:CLASixs}
\end{figure}

In Fig.~\ref{fig:CLASixs} we plot our total cross section results for the
$e p \rightarrow e' p \eta$ process at various $Q^2$ and compare to the
CLAS integrated cross section data~\cite{Thompson:2000by}. The overall
agreement is good, however we notice some deviations in the resonance peak
position. Indeed, the value of the $S_{11}(1535)$ mass extracted by the
CLAS collaboration is 1519~MeV, while 1541~MeV is obtained from our
photoproduction fit. We choose not to change our photoproduction fit value
for these electroproduction data.

\begin{figure}
  \centering
  \includegraphics[width=10.5cm]{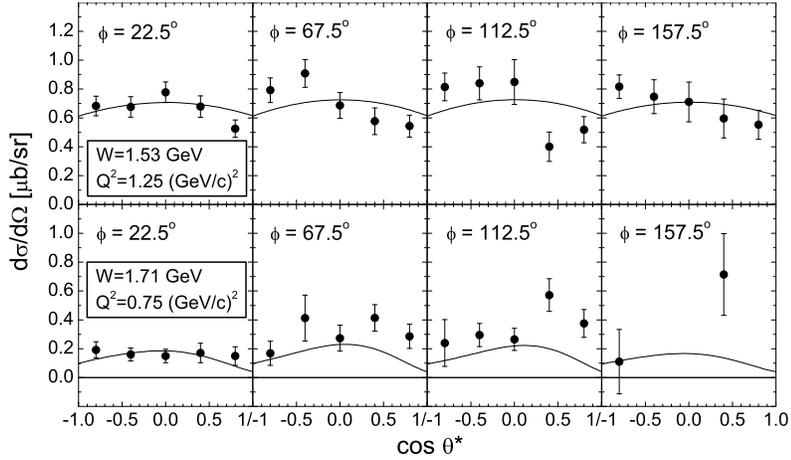}
  \caption{Differential cross section for $p\,(e,e'p)\,\eta$ at various
  $Q^2$ compared to the CLAS data~\cite{Thompson:2000by}.}
  \label{fig:CLASdxs}
\end{figure}

The results for the differential cross sections are shown and compared with
the CLAS data in Fig.~\ref{fig:CLASdxs}. Discrepancies can be seen at some
data points. However, because the statistics of these data is not very
good, our results are still within reasonable error regions of these CLAS
data. More $\eta$ electroproduction data from CLAS are being analyzed and
should be available with better statistics in the near future.

\begin{figure}
  \centering
  \includegraphics[width=6cm]{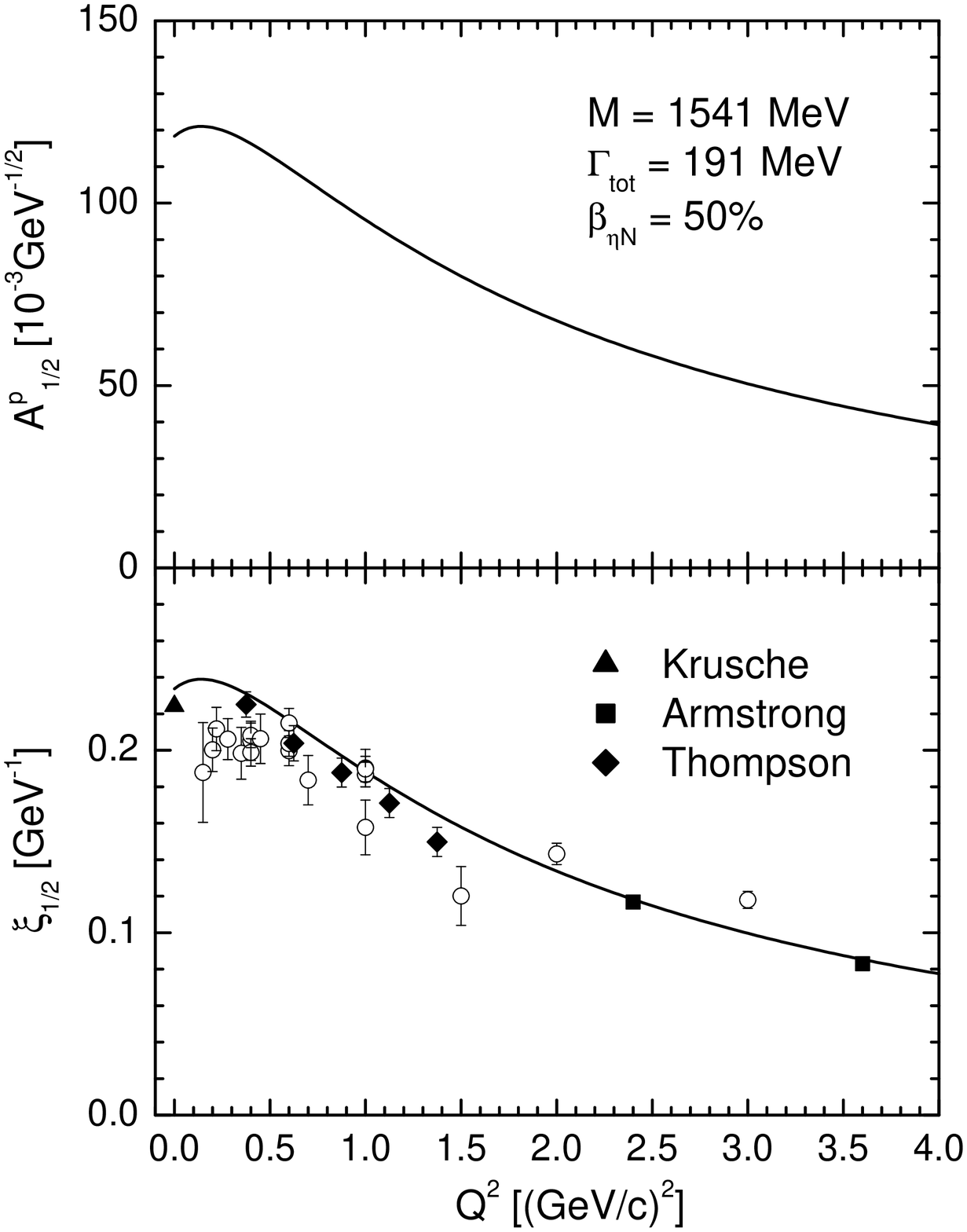}
  \caption{Upper panel: Our results for the photon helicity amplitude
  $A^p_{1/2}(Q^2)$ for $S_{11}(1535) \rightarrow \gamma p$.
  Lower panel: The result for the quantity
  $\xi_{1/2}$ defined in Eq.~(\ref{eq:xi}),
  compared with the values extracted from recent JLab data,
  Armstrong {\it et~al.}~\cite{Armstrong:1998wg} and
  Thompson {\it et~al.}~\cite{Thompson:2000by}, and
  previous data from Ref.~\cite{Brasse}.
  The photoproduction data point ($Q^2=0$) is from
  Krusche {\it et~al.}~\cite{Krusche:1995nv}.}
  \label{fig:S11Q2}
\end{figure}

In Fig.~\ref{fig:S11Q2} we show our result for the $Q^2$ dependence of the
photo-excitation helicity amplitude $A^p_{1/2}(Q^2)$ for $S_{11}(1535)
\rightarrow \gamma p$. The form of the $Q^2$ dependence has been described
in Eq.~(\ref{eq:S11Q2}). In order to avoid large model uncertainties
arising from different values of partial and total widths of the
$S_{11}(1535)$ employed in other analyses, we choose not to compare with
the helicity amplitudes $A^p_{1/2}(Q^2)$ extracted from other analyses.
Instead, we compare the nearly model-independent quantity introduced by
Benmerrouche {\it et al.}~\cite{Benmerrouche:1995uc},
\begin{equation}\label{eq:xi}
  \xi_\lambda = \sqrt{\frac{m_N k_R}{M_R q_R}
  \frac{\beta_{\eta N}}{\Gamma_{tot}}}\, A_\lambda \,,
  \qquad \mbox{for $\lambda$ = 1/2, 3/2} \,.
\end{equation}
As opposed to the uncertainty from $\beta_{\eta N}$ and $\Gamma_{tot}$
between different analyses, the quantity $\xi_\lambda$ is almost
independent of the extraction process. In Fig.~\ref{fig:S11Q2} we compare
our $\xi_{1/2}(Q^2)$ values for $S_{11}(1535)$ with the ones extracted from
the recent JLab data~\cite{Armstrong:1998wg,Thompson:2000by} and older
data~\cite{Brasse}. It is seen that overall good agreement is achieved up
to $Q^2 = 4.0\;\mbox{(GeV/c)}^2$.

\section{Summary and Conclusions} \label{sec:Concl}
In this work we have presented a new study of eta photo- and
electroproduction in an isobar model. The model has been developed
similarly to the unitary isobar model MAID for pion electroproduction. It
is based on nucleon Born terms, vector meson exchange contributions in the
$t$-channel and $s$-channel nucleon resonance excitations described as in
MAID by Breit-Wigner shapes. This allows a direct comparison between our
pion and eta production analyses. Based on experimental data from Mainz,
GRAAL and JLab we have obtained a very good fit of photo- and
electroproduction cross sections and photon asymmetries. We find
significant contributions of the $D_{13}(1520)$, $S_{11}(1535)$,
$S_{11}(1650)$, $D_{15}(1675)$, $F_{15}(1680)$, $D_{13}(1700)$,
$P_{11}(1710)$ and $P_{13}(1720)$ resonances. Generally the resonance decay
widths, the branching ratios and the e.m. helicity couplings are strongly
correlated, and in a photo- or electroproduction experiment only the
combined quantity $\xi$ (see Eq.~(\ref{eq:xi})) can be uniquely determined.
For the $S_{11}(1535)$ the $\eta N$ branching ratio is rather well known
from $(\pi, \eta)$ reactions, giving us the possibility to concentrate our
studies on the photon coupling $A_{1/2}$. Our result of $118 \times 10^{-3}
\;\mbox{GeV}^{-1/2}$ is in good agreement with previous detailed studies
\cite{Krusche:1995nv,Krusche:1997jj}. However, we find a big influence of
the second $S_{11}(1650)$ that is absolutely necessary to describe the
total cross sections at energies above $W=1600\;\mbox{MeV}$. Fitting also
electroproduction data of JLab allowed us to further determine the $Q^2$
evolution of $A_{1/2}(Q^2)$ with a simple parameterization. Our result
confirms the $Q^2$ points of the experimental analyses and yields a form
factor that falls off much more slowly than the nucleon dipole form factor
or the $N\rightarrow \Delta(1232)$ transition form factor.

For the higher resonances we find large branching ratios of $7.9\%$, $17\%$
and $26\%$ for $S_{11}(1650)$, $D_{15}(1675)$ and $P_{11}(1710)$,
respectively, but very small values of only $0.06\%$ for the $D_{13}(1520)$
and $F_{15}(1680)$. A remarkable result is obtained for the $E/M$ ratio of
the $D_{15}(1675)$ that turns out to be only $1.4\%$ (PDG gives $20 \pm 27
\%)$. This ratio can be expressed as
$(A_{1/2}-\frac{1}{\sqrt{2}}A_{3/2})/(A_{1/2}+\sqrt{2}A_{3/2})$ and would
be expected to approach $100\%$ from helicity conservation at large $Q^2$
as in the case of the $N\rightarrow \Delta(1232)$ excitation. An
investigation of this interesting ratio will be possible by an
electroproduction experiment with $R_{TT}$ separation.

Concerning the background processes our fits confirm a very small $\eta NN$
coupling constant of $g_{\eta NN}^2/4\pi=0.10$ with a slight preference for
a pseudoscalar coupling. However, as the coupling strength weakens, this
preference finally diminishes. We have also looked for a possible
``missing" $S_{11}$ resonance in the region around 1700~MeV. However, in
our isobar model such an additional state is not supported by the fits to
the data.

Finally, the only data that we can not sufficiently describe are the target
polarization data in the threshold region. As already found in the model
independent analysis of Ref.~\cite{Tiator:1999gr}, the structure found in
the target polarization goes beyond the scope of an isobar model and would
require a very large ``background" from another mechanism, not considered
in our model, nor in any other model so far. Further measurements of this
and other target polarization observables with double polarization would be
very useful to track down the underlying mechanism of the threshold
$s$-wave production if it should not be given by the $S_{11}(1535)$
resonance alone.

\begin{ack}
The authors would like to thank S.S.~Kamalov and C.~Bennhold for helpful
discussions, and members of the GRAAL and CLAS collaborations for
stimulating this work of our new analysis. W.-T.~C. is grateful to the
Universit\"at Mainz for the hospitality extended to him during his visits.
This work was supported in parts by the National Science Council of ROC
under Grant No.~NSC89-2112-M002-078, by Deutsche Forschungsgemeinschaft
(SFB 443), and by a joint project NSC/DFG TAI-113/10/0.
\end{ack}



\end{document}